\documentclass[prb,aps,twocolumn,superscriptaddress]{revtex4}
\usepackage{graphicx}
\usepackage{amssymb}
\usepackage{bm}

\begin{document}

\author{Toshifumi Itakura}
 \affiliation{ASPRO TEC. 465-0054, Japan}
 \email{itakurat@s6.dion.ne.jp}

\title{The hierarchical spectral flow of Bloch electron}

\begin{abstract}
The Hall conductance of a two-dimensional electron gas has been studied in a uniform magnetic field.
The Hall conductivity is expressed as a sum of two contributions:
one corresponding to the classical Drude-Zener formula,
and a second which has no classical analogy.
The developed theory is applied to the Hall effect.
The Kobo formula is written in a form that makes apparent the quantization
when the Fermi energy lies in a gap.
We examine the hierarchical spectral flow of Hall conductivity.
\end{abstract}


\maketitle

\section{Introduction}
The quantum Hall effect is one of the most spectacular phenomena 
in condensed matter physics.
The integer quantum Hall effect has been observed in two-dimensional
electrons in semiconductor heterostructures,
\cite{Klitzing,Laughlin},
and quasi-one-dimensional organic conductors in the magnetic-filled-induced
spin density wave state.
\cite{FISDW1,FISDW2,FISDW3,FISDW4.FISDW5}
Several recent experiments made \cite{Klitzing,Tsui}
 lead to remarkable conclusion that
the Hall conductivity of a two-dimensional system in its quantum limit is
quantized to integral multiples of $e^2/\hbar$.
However, it can be concluded from Laughlin's
\cite{Laughlin}
argument that the Hall conductance is quantized whenever the Fermi energy
lies in an energy gap, even if the gap lies within a Landau level.
For example, it is known that if the electrons are subject to a weak sinusoidal
perturbation as well as to the uniform magnetic field,
with $\phi=P/Q$ magnetic-flux quanta per unit cell of
perturbing potential, each Landau level is spilt into $Q$ subbands of 
equal weight.
\cite{Bloch}
One might except each of these subbands to give a Hall conductance
equal to $e^2/P h$, and that is what the classical theory of the Hall
current suggests, but according to Laughlin each subband must carry an
integer multiple of the Hall current carried by
the entire Landau level.
This result appears even more paradoxical when it is realized that $Q$,
the number of subbands, can become arbitrarily large
by an arbitrarily small change of the flux density.
The Hall conductance of a two-dimensional electron gas has been studied in a uniform magnetic field.
The Kubo formula is written in a form that makes apparent the quantization
when the Fermi energy lies in a gap.
Explicit expressions have been obtained for the Hall conductance.
\cite{Kohmoto}
A peculiar problem of Bloch electrons in a magnetic field
frequently arises in many different physical contexts.
It resembles some properties of integer Hall effect
\cite{Hofstadter,Azbel} and
is spectrum has an extremely rich structure of Cantor set,
and exhibits a multifractal behavior.
\cite{Kohmoto,Aubry,Thouiess}
And it describes the localization phenomenon in quasiperiodic
potential.
\cite{Hiramoto}
In another context, the hierarchical spectral flow  is examined.
\cite{Weigmann}

\section{Hamiltonian} 

To derive the expression for the conductivity, we shall 
use linear response theory
based on the following assumptions:

(i) The electron system can be described as Fermi-Dirac assembly of
independent quasi particles.

(ii) Only elastic scattering is admissible.

(iii) A two-dimensional solid is formed by a layer in the (x,y) plane
and magnetic field ${\bf B}$ is perpendicular
 to the layer (${\bf B} = ( 0,0,B)$).

(iv) There are gaps in energy spectrum of the one-electron Hamiltonian
describing the system
\begin{equation}
\label{eqm:Hami}
H= (1/2m) [ {\bf P} - (e/c) {\bf A}]^2 + V(x,y)
\end{equation}
where $m$ and $e$  are electron mass and charge respectively, ${\bf A}$
is vector potential (${\bf B} = curl {\bf A}$) and $V(x,y)$ is an
arbitrary flux potential.

(v) An electric field ${\bf E}$ established in the solid results 
in an electric current ${\bf I}$ linearly related to filed through Ohm's law
\begin{equation} 
 {\bf I} = {\bf \sigma} {\bf E}
\end{equation}
where ${\bf \sigma}$ is the conductivity tensor.

\section{Linear response theory}

The diagonal components are given by the following expression:
\cite{Streda}
\begin{equation}
\label{eqn:diagonal}
\sigma_{ii} = \sigma_{ii} ( E_f,0) = \pi \hbar e^2 {\rm Tr} [v_i \delta ( E_F -H) v_i \delta (E_F - H)]
\end{equation}
and non-diagonal components by
\begin{eqnarray}
\label{eqn:cond0}
\sigma_{ij} &=& \sigma (E_F,0) 
=e^2 \int_{-\infty}^{E_F} A_{ij} ( \eta) d \eta \\
\label{eqn:acond}
A_{ij} ( \eta) &=& i \hbar {\rm Tr} [ v_i ( \frac{d G^+}{d \eta}v_j \delta ( \eta- H) \nonumber \\
&-& v_i \delta ( \eta -H) v_j \frac{d G^-}{d \eta}]
\end{eqnarray}
where the Green function is defined by
\begin{equation}
G^{\pm} ( \eta) = ( \eta -H \pm i 0)^{-1},
 \delta ( \eta -H) = - \frac{1}{2 \pi i} ( G^+ -G^-)
\end{equation}
and velocity operator is given by commutation relation
\begin{equation}
 \label{eqn:comm}
v_i = \frac{1}{i \hbar} [r_i,H]=-\frac{1}{i \hbar} [ r_i, G^{-1}] = \frac{1}{m} ( p_i - \frac{e}{c} A_i)
\end{equation}
To proceed further we split the Hall conductivity into two parts using an expression originally derived 
\cite{Smrcka}
\begin{eqnarray}
\label{eqn:acond1}
 A_{ij} ( \eta) &=& \frac{1}{2} \frac{d}{d \eta} B_{ij} ( \eta ) \nonumber \\ 
&+& \frac{1}{2} {\rm Tr} \frac{ d \delta ( \eta - H)}{d \eta} ( r_i v_j - r_j v_i)\\
B_{ij}( \eta) &=& i \hbar {\rm Tr}[v_i G^+ ( \eta) v_j \delta ( \eta - H) 
\nonumber \\
&-& v_i \delta ( \eta - H) v_j G^- ( \eta)].
\end{eqnarray}
The second term on the right-hand side of equation (\ref{eqn:comm}) can be rewritten into the more convenient form
\begin{equation}
 \label{eqn:cond}
\frac{1}{2} {\rm Tr} \frac{d \delta ( \eta - H)}{d \eta} ( r_x v_y - r_y v_x) = \frac{c}{e} \frac{\partial}{\partial B} {\rm Tr} ( \eta - H)
\end{equation}
if the definition of the Hamiltonian
(\ref{eqn:Hami}), the expression of (\ref{eqn:acond1})  and commutation relation
\begin{equation}
[r_x,v_y]= [r_y,v_x]=0
\end{equation}
are used.
The expression (\ref{eqn:cond}) is valid for arbitrary choices of vector potential $A$, nevertheless the simplest derivation is obtained if the circular gauge centered at the origin is used; $A= \frac{1}{2} B ( -y,x,0)$. Introducing expression (\ref{eqn:acond}), (\ref{eqn:acond1}) and (\ref{eqn:cond}) into (\ref{eqn:cond0}) we get immediately useful expression for the Hall conductivity
\begin{eqnarray}
\label{eqn:offdiagonal}
\sigma_{ij} &=& \sigma_{ij}^{I} + \sigma_{ij}^{II}  \nonumber  \\
\sigma_{ij}^I &=& \frac{e^2}{2} i \hbar {\rm Tr} [v_i G^{+} ( E_F) v_j \delta ( E_F - H) \nonumber \\
&-&v_i \delta ( E_F - H) v_j G^- ( E_F)]  \nonumber \\
\sigma_{xy}^{II} &=& - \sigma_{yx}^{II} = ec \frac{\partial N (E)}{\partial B}|_{E=E_F}
\end{eqnarray}
where $N(E)$ is the number of states below the energy E defined by
\begin{equation} 
N(E) = \int_{-\infty}^E {\rm Tr} \delta ( \eta - H) d \eta
\end{equation}
and the derivative with respect to magnetic field B is taken at the Fermi energy.
The first term $\sigma_{xy}^{I}$ depends on the structure of solid, crystallographic orientation and of course on the potential $V(r)$.
The classical Drude-Zener result if vertex corrections are omitted is given by
\begin{equation}
\sigma_{xy}^I = - \omega \tau \sigma_{xx}
\end{equation}
where $ \omega = |e| B/mc$ is cyclotron frequency; the lifetime $ \tau$ is equal to
$\hbar/2\Gamma$ $(\Gamma= - {\rm Im} \Sigma( E_F) )$.

\section{Hall conductivity}

To derive the expression for the quantized Hall conductivity,
we shall employ the assumption (iv) mentioned above,
namely that there are gaps in electron energy spectrum in magnetic field
 and that the Fermi energy is lying just within a gap,
where the density of  states is zero.
Since $\delta$ functions in expressions (\ref{eqn:diagonal}) 
and (\ref{eqn:offdiagonal}) describe contributions to the density at
Fermi energy, diagonal elements of the conductivity tensor
$\sigma$ and classical term of the Hall conductivity $\sigma^I_{xy}$ are equal
to zero and we get
\begin{equation}
\label{eqn:topo}
\sigma_{xx}=\sigma_{yy}=0, 
\sigma_{xy}= - \sigma_{yx} = ec ( \frac{\partial N(E)}{\partial B})|_{E=E_F}
\end{equation}
We define band conductivity 
\begin{equation}
\sigma_k = \frac{ \delta N_k}{ \delta \eta}
\end{equation}
where $\eta= \frac{P}{Q}$ and $N_k$ is defined by
\begin{equation}
N_{k} = N_{k-1} + p_k N(k)
\end{equation}
where $N(k)$ is the number of state in $k$-th band and $k=1,...Q$.
and $p_k=\pm$ corresponds to electron or hall. Using the equation,
\begin{equation}
\frac{ \delta N_k}{\delta \eta}
 = \frac{\delta N_{k-1}}{\delta \eta} + p_k \frac{\delta N(k)}{\delta \eta}
\end{equation}
The equation of Hall conductivity is given by
\begin{equation}
\sigma_k = \sigma_{k-1} + p_k \sigma (k)
\end{equation}
where $\sigma(k)$ is $k$-th band Hall conductivity.
We need the notion of strings, a Hall conductance,
a hierarchical spectral flow, and a hierarchical tree.
\cite{Wiegmann}
The Hall conductance $\sigma(k)$ of the band k=$1,...Q$
is the Chern class of the band, the number of
zeros of the wave function as a function of $k_x,k_y$ in the Brillion
zone.
We also use the conductance of $k$ filled bands $\sigma_k$.
The latter is assigned to the $k$th gap.
They are determined by the Diophantine equation,
\cite{Kohmoto}
\begin{equation}
P \sigma_k = k ({\rm mod} Q)
\end{equation}
restricted to the range $-Q/2 < \sigma_k \leq Q/2$.

Regarding the hierarchical spectral flow,
we refer to the band k' of the problem with period $\eta' = P'/Q'$
as the parent band of the parent generation corresponding to the daughter
band k of the generation $\eta=P/Q$ if 
the Hall conductance of the band $k$ is the ratio of the difference of the
number of states of parent and daughter bands to the difference of the periods
of generation.

We next use following properties,
\begin{eqnarray}
a &=& b q + p_1 r, 0<r<b, \\
b &=& r q_1 + p_2 r_1, 0<r_1<r, \\
r &=& r_1 q_2 + p_3 r_2, 0<r_2<r_1,\\
... &&...................\\
r_m &=& r_m q_{m+1}, 0<r_m < r_{m-1}
\end{eqnarray}
Therefore, we rewritten the equations,
\begin{eqnarray}
\frac{a}{b} &=& q + p_1 \frac{r}{b}, \\
\frac{b}{r} &=& q_1+ p_2 \frac{r_1}{r},  \\
\frac{r}{r_1} &=& q_2  + p_3 \frac{r_2}{r_1}, \\
....&&...........\\
\frac{r_{m-1}}{r_m} &=& p_{m+1} q_{m+1}, 
\end{eqnarray}
Thus,
\begin{equation}
\frac{a}{b}=q+\frac{p_1}{q_1 +  
\frac{p_1}{q_2 + \frac{p_2}{q_3  + --- \frac{p_{Q+1}}{q_{Q+1}}}}}
\end{equation}
%Due to the above relation,
%\begin{equation}
%\frac{r_{k-1}}{r_k} = \sigma_k, 
%q_k = [ \frac{r_{k-1}}{r_k} ]=[\sigma_k ]=\sigma(k).
%\end{equation}
Thus
\begin{eqnarray}
\sigma_0 &=& \frac{Q}{P} = \frac{a}{b}, \nonumber \\
\sigma_k&=&\frac{p_k}{q_{k+1} +
 \frac{p_{k+1}}{q_{k+2} + \frac{p_{k+2}}{q_{k+3} --- 
+\frac{p_{Q+1}}{q_{Q+1}}}}} 
\end{eqnarray}
$\sigma_0$ is first brillouin zone conductivity, thus given by $\frac{Q}{P}$.
These equations express the conductance of Bloch electron system in strong magnetic field.
The hierarchical spectral flow is this fractional expression of Hall 
conductivity.

\section{Discussion}
We have examine the Hall conductance of Hofstadter Butterfly.
The hierarchical spectral flow is this fractional expression of Hall 
conductivity.

We limit ourselves to rational magnetic field, namely
\begin{equation}
\frac{|e|}{h c} {\bf B} \cdot {\bf a}_1 \times {\bf a}_2 
\times \cdots \times {\bf a}_k,
\end{equation}
where $k=1,\cdots,Q$
Let us suppose that $\sigma (k)$ narrow bans are lying below $E_F$.
Since the magnetic field is supposed to be perpendicular the two-dimensional layer,
the number of state is given by
\begin{equation}
\label{eqn:num}
N=\sigma (k) \frac{1}{J^2 | {\bf a}_1 \times {\bf a}_2 
\times \cdots \times {\bf a}_k |} J = \sigma (k) \frac{|e|}{hc} B.
\end{equation}
At the bottom of the band, where broad gaps exist, the small changes in magnetic field
do not change the number of Landau levels below Fermi energy and we get
\begin{equation}
\sigma_{xy} = - \frac{e^2}{h} \sigma (k).
\end{equation}
using expression (\ref{eqn:topo}) and (\ref{eqn:num}).
The Hall conductance $\sigma(k)$ of the band k=$1,...Q$
is the Chern number
\cite{Goldman}
of the band.
This is the number of
zeros of the wave function as a function of $k_x,k_y$ in the Brillion
zone.

%----------------------

\end{document}